\documentclass[letterpaper]{ptephy_v1}

\begin{document}

\title{Probing the string axiverse by gravitational waves from Cygnus X-1}


\author[1]{Hirotaka Yoshino}
\affil{Institute of Particle and Nuclear Studies,
KEK, Tsukuba, Ibaraki, 305-0801, Japan \email{hyoshino@post.kek.jp}}

\author[1,2]{Hideo Kodama}
\affil{Department of Particle and Nuclear Physics,
Graduate University for Advanced Studies, Tsukuba 305-0801, Japan}


\begin{abstract}%
In the axiverse scenario, a massive scalar field (string axion) 
forms a cloud around a rotating black hole (BH) by superradiant 
instability and emits continuous gravitational waves (GWs). 
We examine constraints on the string axion parameters that can 
be obtained from GW observations. If no signal is detected in a 
targeted search for GWs from Cygnus X-1 in the LIGO data
taking account of axion nonlinear self-interaction effects, the decay 
constant $f_a$ must be smaller than the GUT scale in the mass range 
$1.1\times 10^{-12}\mathrm{eV}<\mu<2.5\times 10^{-12}\mathrm{eV}$. 
Possibility of observing GWs from 
invisible isolated BHs is briefly discussed.
\end{abstract}

\subjectindex{E31, E02, C15, B29}

\maketitle

%
%
\section{Introduction}

The second-generation ground-based gravitational wave (GW) detectors 
will begin operations within a few years and provide us with a new eye 
to discover various new phenomena, which include those caused 
by fundamental fields in hidden sector.  Promising candidates 
for such hidden sector objects are string axions with 
tiny masses~\cite{Arvanitaki:2009,Arvanitaki:2010,Kodama:2011}.
In string theories, a variety of moduli appear when the extra dimensions 
get compactified, and perturbatively, some of them are predicted 
to behave as massless pseudo-scalar fields due to shift symmetry 
from the four-dimensional perspective. Because of nonperturbative effects, 
these massless fields are expected to acquire small masses. 
The axion masses are naturally expected to be uniformly
distributed in the logarithmic scale
in the range $-33\lesssim \log_{10}(\mu\mathrm{[eV]})\lesssim -10$. 
When their 
Compton wavelengths are astrophysical scales, they may cause new 
astrophysical phenomena that can be observed by GWs.

Suppose the low-energy effective theory contains a string axion 
with mass $\mu$. Then, around a rotating black hole (BH) with mass $M$, 
the axion field is known to extract the BH rotation energy through 
the superradiant instability and forms an axion cloud around the BH, 
if $M\mu$ is $O(1)$ in the natural units $c=G=\hbar=1$. 
Such an axion cloud causes rich phenomena due to its self-interaction, 
and also emits GWs~\cite{Yoshino:2012,Yoshino:2013}.
In particular, it always emits continuous GWs with frequency 
$\tilde{\omega}\approx 2\mu$.

Searches for continuous GWs have been already done for the data of 
the LIGO science runs, assuming that their sources are rotating 
distorted neutron stars (see \cite{Aasi:2012} for a recent report 
and references for other searches). 
An important feature of the continuous wave search is that 
sensitivity can be improved with the increase in the observation time 
$T_{\rm obs}$ because the lower limit on detectable GW amplitudes 
is given by $h_0\sim O(100)\sqrt{S_n/T_{\rm obs}}$ with $S_n$ 
being the noise spectral density, when the frequency width of the 
GW is smaller than $1/T_{\rm obs}$. Utilizing this feature, 
the LIGO team derived the strong upper limit on the amplitude, 
$h_{\rm UL}\sim 10^{-24}$, from the null detection in the observation data 
of order one year.

The purpose of the present paper is to point out that the same method 
can be applied to continuous GWs from the well-known stellar-mass BH 
in Cygnus X-1 to obtain definite constraints on string axion parameters 
by the LIGO data and the future data from the second-generation detectors. 
We also discuss the possibility to apply a similar idea 
to nearby invisible isolated BHs.

%
%
\section{BH-axion system}

In this paper, the field $\Phi$ is assumed to be real and 
to obey the Sine-Gordon equation,
\begin{equation}
\nabla^2\varphi - \mu^2\sin\varphi = 0,
\label{Eq:Sine-Gordon}
\end{equation}
where $\varphi:=\Phi/f_a$ is the amplitude normalized 
by the decay constant $f_a$.
The potential term in this equation naturally arises 
by the nonperturbative instanton effect for the QCD axion, 
and a similar mechanism is expected for 
string axions~\cite{Svrcek&Witten:2006}. Although $f_a$ and $\mu$ 
are related to each other in the QCD axion case, they are treated 
as independent parameters for string axions. When $\varphi$ is small, 
the Sine-Gordon equation is well approximated by the Klein-Gordon equation, 
$\nabla^2\varphi - \mu^2\varphi = 0$, while
the nonlinear effect becomes relevant for $\varphi\sim 1$.

Quasibound states of the Klein-Gordon field around a Kerr BH have been well studied~\cite{Detweiler:1980,Zouros:1979,Furuhashi:2004,Cardoso:2005,Dolan:2007}.
Because there is an ingoing flux across the horizon, 
each eigenfrequency takes a complex value,
\begin{equation}
\omega = \omega_R+i\omega_I.
\end{equation}
If the discrete real part $\omega_R$ satisfies the superradiant 
condition $\omega_R <m\Omega_H$, where $m$ is the azimuthal quantum 
number and $\Omega_H$ is the angular velocity of the horizon, 
the energy flux across the horizon becomes negative and $\omega_I$ 
becomes positive. This indicates that the scalar field amplitude 
grows exponentially.
The typical time scale of this {\it superradiant instability} is 
$T_{\rm SR}\gtrsim 10^{7}M$, which is around one minute for a solar-mass BH. 
In the case of $M\mu\ll 1$, eigenstates can be obtained by the method 
of matched asymptotic expansion~\cite{Detweiler:1980}. In this approximation,
a solution for $\Phi$ in a distant region is obtained 
from a solution to the nonrelativistic Schr\"odinger equation 
for the hydrogen atom by replacing $e^2$ with $M\mu$,
and each mode is labeled by the angular quantum numbers $\ell$ and $m$
with $-m\le \ell\le m$
and the principal quantum number $n$. Here,
the principal quantum number is defined as
$n=\ell+1+n_r$ in terms of the radial quantum number $n_r=0,1,2,...$
that 
characterizes the oscillatory behavior of the mode function in the radial direction.
The unstable mode function 
with $\ell = m = 1$  and $n=2$ reads
\begin{equation}
\Phi \approx  (M\mu)^2\sqrt{\frac{E_a}{8\pi M}}(kr)e^{-kr}\sin\theta\cos(\omega t-\phi),
\label{L1M1-wavefunction}
\end{equation}
where $E_a$ is the total energy of the axion cloud and $k:=M\mu^2/2$.
The angular frequency for this state is $\omega \approx \mu [1-(M\mu)^2/8]$, 
and hence, $\omega\approx \mu$ holds for $M\mu\ll 1$.

%
\begin{figure}[tb]
\centering
\includegraphics[width=0.6\textwidth,bb=0 0 1024 768]{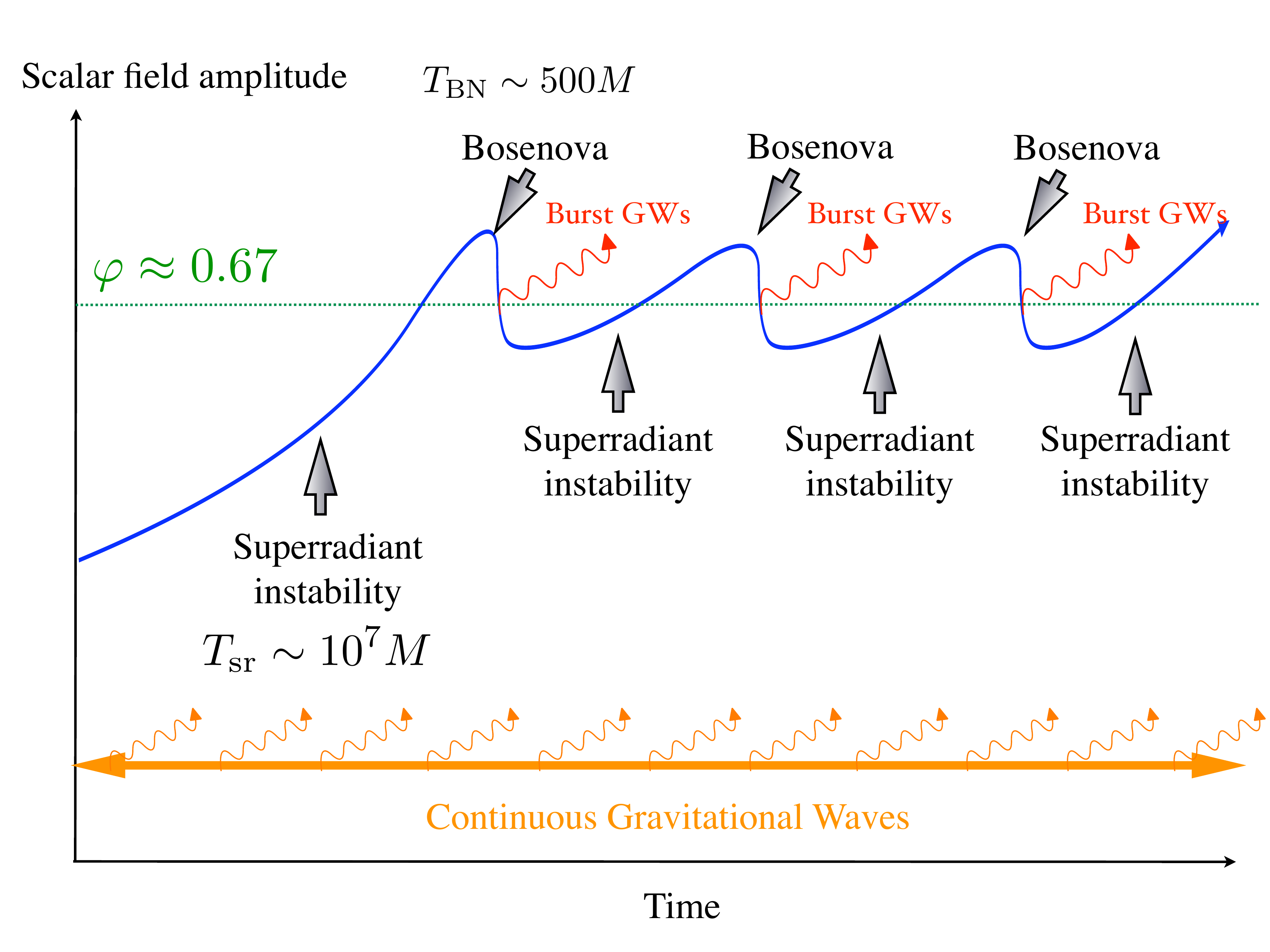}
\caption{
Schematic picture for time evolution of the 
scalar field amplitude and emitted GWs. See text for details. 
}
\label{schematic-figure}
\end{figure}
%

As $\varphi$ becomes larger, the nonlinear effect becomes important. 
In our previous paper \cite{Yoshino:2012}, we studied this phase 
by numerical simulations of the axion cloud 
in the $\ell = m = 1$ mode with $M\mu=0.4$. 
At some point with $\varphi\approx 0.67$, a new mode is suddenly excited, and it carries
positive energy to the horizon and to the far region  
terminating the superradiant instability. 
We call this phenomenon ``bosenova''. The typical time scale of 
the bosenova is $\sim 500M$, and about $5\%$ of the axion cloud energy 
falls into the BH. Then, the system again settles to the superradiant phase. 
In a long time simulation, the system was observed to alternate 
between the bosenova and the superradiant phase. A schematic picture
for the time evolution
of the field amplitude is shown in Fig.~\ref{schematic-figure}.

Burst GWs are generated by the infall of the axion cloud 
energy during the bosenova. 
In our order estimate~\cite{Yoshino:2012},
the GW frequency is within the observation bands of the 
ground-based detectors, but its amplitude may be marginal to be detected 
by the second-generation detectors in the case an axion field with 
the decay constant $f_a\approx 10^{16}~\mathrm{GeV}$ causes a bosenova 
at Cygnus X-1.

In addition to burst GWs from bosenovae, an axion cloud continuously emits GWs
by the level transition of axions and the two-axion annihilation \cite{Arvanitaki:2010}.
The former process can be calculated by the quadrupole approximation~\cite{Arvanitaki:2010}, while the latter process requires direct calculations of a perturbation equation \cite{Yoshino:2013}.
Among these two, the two-axion annihilation is the primary process,
and we discuss its observational consequence in this paper.
In this process, the energy-momentum tensor $T_{\mu\nu}$ fluctuating with the angular frequency $2\omega$ generates monochromatic GWs with the same frequency
\begin{equation}
\tilde{\omega}\approx 2\mu.
\label{omegaGW}
\end{equation}
From an axion cloud in the $\ell = m = 1$ mode, 
GWs in the $\tilde{\ell}=\tilde{m}=2$
mode are radiated. In Ref.~\cite{Yoshino:2013}, 
we found the approximate formula for $M\mu\ll 1$ 
by solving the perturbation equation of a flat background spacetime,
\begin{equation}
h_0\approx \sqrt{\frac{5C_{n\ell}}{2}}
\left(\frac{E_a}{M}\right)(M\mu)^6\left(\frac{M}{d}\right),
\label{GW-amplitude-1}
\end{equation}
where $d$ is the distance to the BH. Here, 
the functional form of Eq.~\eqref{GW-amplitude-1} is reliable 
except for a factor (the value of $C_{n\ell}$ cannot be determined 
within this approximation). We also directly calculated the 
GW radiation rate numerically for a Kerr background, 
and checked that Eq.~\eqref{GW-amplitude-1} holds 
with $C_{n\ell} \approx 10^{-2}$.
Our conclusion of Ref.~\cite{Yoshino:2013} is that the energy 
loss rate by the GW radiation is smaller than the energy extraction 
rate of the axion cloud, and hence, the axion cloud grows until the 
bosenova happens. 
Note that we have ignored the nonlinear self-interaction effects
and used the solution for the quasibound state of the linear Klein-Gordon 
field in estimating the GW amplitude \eqref{GW-amplitude-1}. 
This is a rather strong approximation, and we will come back to this point
later.

Since continuous waves from a distorted neutron 
star are also 
in the $\tilde{\ell} = \tilde{m} = 2$ mode 
in the quadrupole approximation, 
GWs from the $\ell = m = 1$ axion cloud share the
similar features (the angular pattern and the ratio between 
the plus and cross modes) 
with GWs from a neutron star. Therefore, the same method of 
the continuous wave search can be applied to GWs from axion clouds.

%
%
\section{Method for constraining string axion models.}

The frequency region where continuous waves have been analyzed 
is $50~\mathrm{Hz}\le f\le 1200~\mathrm{Hz}$ \cite{Aasi:2012}. 
Since the angular frequency of continuous waves from an axion cloud 
is related to its mass through Eq.~\eqref{omegaGW}, 
we consider the corresponding axion mass range 
\begin{equation}
1.0\times 10^{-13}\mathrm{eV} \le\mu\le 2.5\times 10^{-12}\mathrm{eV}.
\end{equation}
This covers a certain range of the possible mass values of string axions.
For the BH mass $M\approx 15M_\odot$ to be considered in this paper, 
the parameter $M\mu$ is in the range $0.0125\le M\mu\le 0.3$ and is 
relatively small. In this parameter range, the fastest unstable mode 
is the $\ell = m = 1$ mode with $n=2$. Although other unstable modes 
may also grow later, we ignore their contribution because the GW 
emission from these modes is much smaller \cite{Yoshino:2013}. 
Then, we can use the approximate formula for the emitted GW amplitude, 
Eq.~\eqref{GW-amplitude-1}.

We determine the value of $E_a/M$ as follows. As mentioned in the previous section, the superradiant instability of an axion cloud is saturated around $\varphi:=\Phi/f_a\approx 0.67$.
Therefore, the energy content in this situation is given by the formula
\begin{equation}
\varphi_{\rm max} \approx
\frac{\exp(-1)}{\sqrt{8\pi}}\sqrt{\frac{E_a}{M}}
\left(\frac{f_a}{M_{\rm pl}}\right)^{-1}(\mu M)^2\approx 0.67.
\label{Amplitude-Saturate}
\end{equation}
Substituting $E_a/M$ determined by this equation into Eq.~\eqref{GW-amplitude-1}, we derive the condition
\begin{equation}
h_0\approx 1.2\times 10^{-22}\left(\frac{f_a}{10^{16}~\mathrm{GeV}}\right)^2
\left(\frac{\mu}{10^{-12}~\mathrm{eV}}\right)^2
\left(\frac{M}{15M_{\odot}}\right)^3
\left(\frac{d}{1~\mathrm{kpc}}\right)^{-1}
<h_{\rm UL}.
\label{Condition-Amplitude}
\end{equation}
Here, the left-hand side is the amplitude expressed in the axion parameters $(\mu, f_a)$ and the BH
parameters $(d,M)$, and $h_{\rm UL}$ on the right-hand side is the upper bound on the GW amplitude derived from observations. Fixing the BH parameters $(d,M)$, this inequality gives a
constraint on the axion parameters $(\mu, f_a)$.

Before applying the above argument to Cygnus X-1,
we note some subtleties. 
Cygnus X-1 is known to have a large spin parameter, 
$a/M\gtrsim 0.983$, assuming the accretion disk model 
\cite{Reid:2011,Orosz:2011,Gou:2011,Gou:2014}.
If Cygnus X-1 wears an axion cloud, 
the BH interacts with both the accretion disk and the
axion cloud, gradually changing $M$ and $J$.
Therefore, the consistency with the observed spin parameter must be 
checked. 
Recently, the adiabatic evolution of the BH 
parameters was studied for a system of a BH 
wearing a Klein-Gordon field 
(without nonlinear self-interaction) \cite{Brito:2014}.
Initially, the accretion disk spins up the 
BH by supplying angular momentum
\cite{Bardeen:1970,Thorne:1974,Gammie:2003,Benson:2008}
if the BH mass is small, $\mu M\ll 1$.
When $\mu M$ becomes important, the scalar field extracts the
BH angular momentum and the spin parameter $a/M$ drops
until the superradiant condition becomes
marginally satisfied, $\mu\approx m\Omega_H$.
After that, the spin parameter gradually
increases to unity approximately keeping the marginal superradiant
condition. Here, we point out that 
the evolution depends on the
value of the decay constant $f_a$ if the
nonlinear self-interaction is present, because 
the bosenova occurs and the growth of the axion cloud
effectively stops when $\Phi\approx f_a$ (Fig.~\ref{schematic-figure}).
If $f_a$ is order
of or smaller than the
GUT scale, the bosenova typically happens much 
before the axion cloud significantly decreases the spin
parameter: See Ref.~\cite{Yoshino:2012} and
condition~\eqref{Condition-Gravity} below.
For this reason, we assume that the axion cloud scarcely
decreases the BH spin parameter and 
the high spin parameter of Cygnus X-1 does not
contradict the existence of the axion cloud.

Another subtlety is that although we have treated the scalar field 
as a test field in Refs.~\cite{Yoshino:2012,Yoshino:2013}, 
its gravity becomes strong as the axion total energy is increased. 
In Ref.~\cite{Brito:2014}, it was argued that the gravitational backreaction
is not important for a small $M\mu$ because
the axion cloud spreads over a large scale. 
But since there remains a possibility 
that the gravity of an axion cloud may affect the estimate on the
BH parameters by changing the properties of the
accretion disk, we adopt the region where ${E_a}/{M}<0.05$ 
is satisfied.  
Using Eq.~\eqref{Amplitude-Saturate}, this criterion 
can be expressed as
\begin{equation}
\left(\frac{f_a}{10^{16}~\mathrm{GeV}}\right)
<
0.44\times
\left(\frac{M}{15M_\odot}\right)^2
\left(\frac{\mu}{10^{-12}~\mathrm{eV}}\right)^2.
\label{Condition-Gravity}
\end{equation}
%

%
%
\section{Expected constraints from Cygnus X-1}

Now we apply the above argument to Cygnus X-1.
The Cygnus X-1 is in binary with a companion star, and 
accretion of matter from a companion star makes it possible to 
observe the phenomena around the BH. 
The recent observation \cite{Reid:2011,Orosz:2011,Gou:2011,Gou:2014}
determines the distance from the earth, the mass and the spin parameter as 
$d=1.86^{+0.12}_{-0.11}~\mathrm{kpc}$, $M=14.8\pm 1.0M_{\odot}$,
and $a/M\gtrsim 0.983$. 
The inclination angle of the orbital plane is $i = 27.1\pm 0.8~\mathrm{deg}$.
Substituting $d=1.86~\mathrm{kpc}$ and 
$M=15M_\odot$ into the inequality~\eqref{Condition-Amplitude}, we have
\begin{equation}
6.3\times 10^{-23} \left(\frac{f_a}{10^{16}\mathrm{GeV}}\right)^2
\left(\frac{\mu}{10^{-12}\mathrm{eV}}\right)^2< h_{\rm UL}.
\label{Condition1-CygX1}
\end{equation}
%

%
\begin{figure}[tb]
\centering
\includegraphics[width=0.5\textwidth,bb=0 0 360 381]{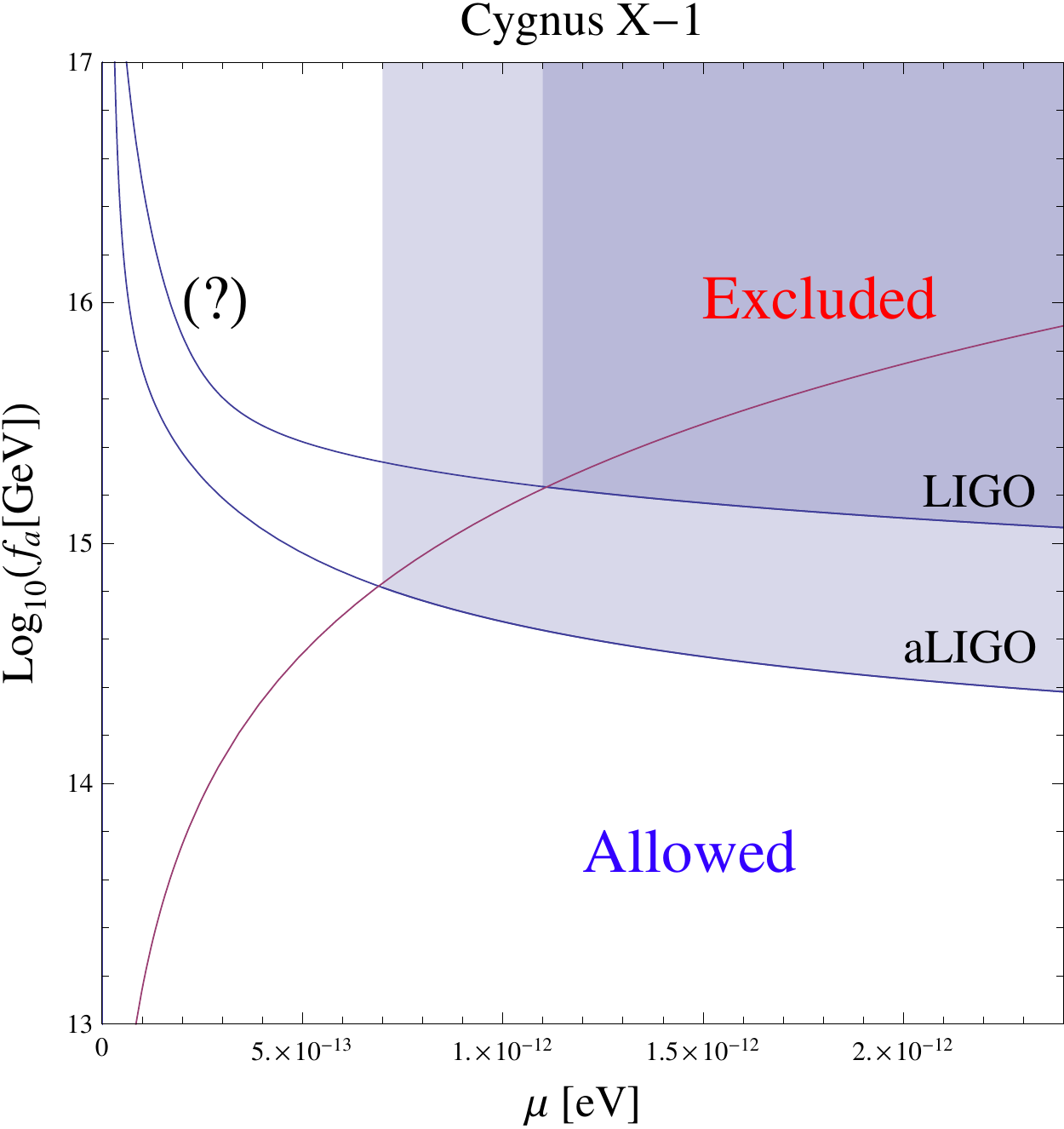}
\caption{Expected constraints for the mass $\mu$ and the decay constant
$f_a$ of string axions derived by continuous GWs from Cygnus X-1.
The cases for the LIGO and aLIGO detectors are shown.
Also shown is the curve above which the gravity effect of the axion cloud
becomes important. The constraints may not be reliable
above this line as indicated by ``(?)''.
}
\label{mass-fa-constraint-CygX1-aLIGO}
\end{figure}
%

Figure~\ref{mass-fa-constraint-CygX1-aLIGO} shows the expected 
constraints in the parameter space $(\mu, f_a)$
that come from the observations
by the LIGO and the Advanced LIGO (aLIGO).
The upper one of the two monotonically decreasing curves is 
the border line of the inequality~\eqref{Condition1-CygX1} 
for the LIGO observation. 
Here, we have substituted the upper limit
on the continuous wave amplitude derived 
by LIGO's all-sky search~\cite{Aasi:2012} into $h_{\rm UL}$. 
Note that the constraint given in this way must be
interpreted as a theoretical forecast, 
because the authors of \cite{Aasi:2012} looked 
for continuous waves from isolated neutron stars and their result cannot 
be applied to GWs from binaries like Cygnus X-1 in which
the binary motion causes the frequency modulations by the Doppler
shift. In order to obtain a reliable value 
for $h_{\rm UL}$,
a targeted search with matched filtering 
for GWs from Cygnus X-1
has to be carried out.
The border line of the criteria \eqref{Condition-Gravity} 
is depicted by the monotonically increasing curve, above which 
the gravity effect of the axion cloud may become important. 
These two curves intersect at $\mu\approx 1.1\times 10^{-12}\mathrm{eV}$, 
and therefore, the border line of the condition \eqref{Condition1-CygX1} 
is not reliable in the region $\mu< 1.1\times 10^{-12}\mathrm{eV}$ 
as indicated by ``(?)''. In contrast, for 
$\mu> 1.1\times 10^{-12}\mathrm{eV}$, the parameters on the border line 
satisfy the condition \eqref{Condition-Gravity} 
and the curve is reliable. Since the GW amplitude is 
expected to be a monotonically increasing
function of $f_a$ for a fixed $\mu$, we exclude all of the region 
above this border line. In particular, the decay constant 
$f_a\approx 10^{16}\mathrm{GeV}$, which seems one of the natural 
choices~\cite{Arvanitaki:2009}, is excluded in the mass 
range $1.1\times 10^{-12}\mathrm{eV}<\mu< 2.5\times 10^{-12}\mathrm{eV}$.

The lower one of the two decreasing curves is 
for the aLIGO observation 
(the curves from the other second-generation detectors are similar). 
Here, we assumed $h_{\rm UL}$ 
to be given by $\approx 150\sqrt{{S_n}/{T_{\rm obs}}}$ 
with $\sqrt{S_n}$ the design sensitivity presented 
in \cite{Harry:2010} and the observation time $T_{\rm obs}=5000$ hours. 
The two curves intersect at $\mu\approx 0.7\times 10^{-12}\mathrm{eV}$. 
Since the sensitivity of the aLIGO detector is 10 times better than 
that of the LIGO detector, the constraint on $f_a$ can be 
improved by a factor of three. 
In particular, the value of $f_a$ of $0.1\times\textrm{GUT scale}$ 
is excluded in the range 
$0.7\times 10^{-12}\lesssim\mu\lesssim 2.5\times 10^{-12}$.

On the other hand, if continuous GWs from Cygnus X-1 are detected, 
we can determine the values of $(\mu, f_a)$. 
Note that continuous GWs from Cygnus X-1 can be clearly 
distinguished from other continuous GWs from distorted neutron stars
for the following reason. Due to the Doppler shift
caused by the binary motion, continuous GWs from Cygnus X-1 has
frequency modulation  with the same period as the orbital period.
On the other hand, continuous GWs from distorted neutron stars
do not have such frequency modulation if they are isolated,
and have different periods of frequency modulation if they are in binary. 
Therefore, 
the information  unique to Cygnus X-1 
is encoded in the wave forms. Since the sensitivity to
signals of continuous GWs highly depends on the phase behavior,
the distinction is possible if a targeted search is carred out
taking account of 
such frequency modulation.

%
\begin{figure}[tb]
\centering
\includegraphics[width=0.7\textwidth,bb=0 0 360 135]{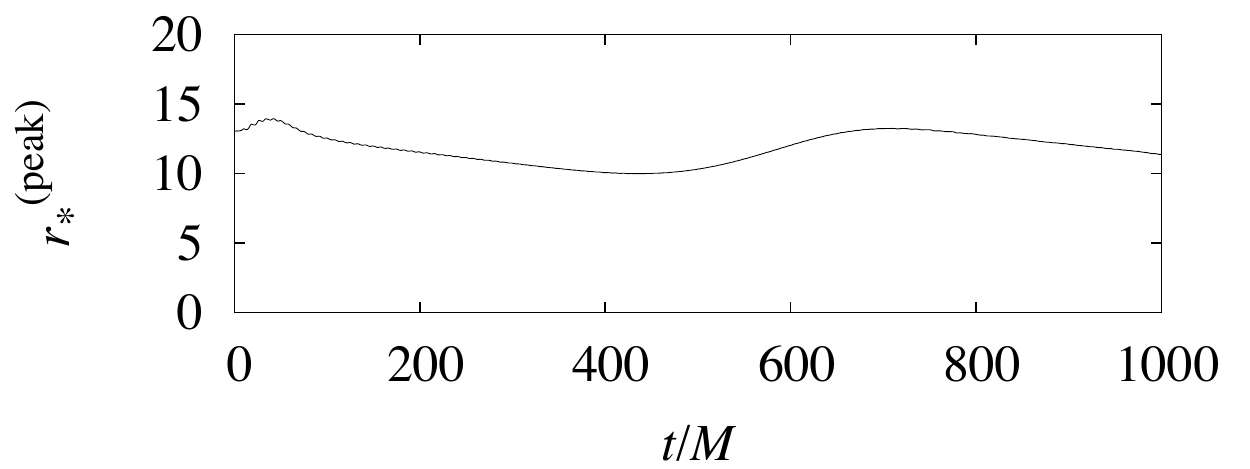}
\caption{The time dependence of the radial position 
(in the tortoise coordinate $r_*$) of the field peak of the axion cloud
in the simulation for $M\mu=0.4$ with the initial amplitude $\varphi=0.60$.
The axion cloud oscillates in the radial direction.  
This figure is taken from Ref.~\cite{Yoshino:2012}.
}
\label{radial-oscillation}
\end{figure}
%

A potential problem of the discussion here is that 
we have ignored the nonlinear self-interaction effects
and assumed the axion cloud to radiate purely monochromatic waves. 
In reality, the self-interaction causes a relatively complicated dynamics. 
Figure \ref{radial-oscillation} shows the time evolution of the 
peak position of the axion cloud in the case of $M\mu=0.4$ with the initial
amplitude $\varphi\approx 0.60$ 
in our simulation~\cite{Yoshino:2012}.
Since the axion cloud oscillates in the radial direction,
the GW frequency is expected to modulate by a factor of few \% due to
the gravitational redshift effect. 
This point requires a careful check by direct numerical calculations
of GWs in the presence of axion nonlinear self-interaction. 
If the frequency modulation with the amplitude
$\Delta\omega$ is present, quadratic estimators cannot improve
the signal-to-noise ratio even if we make the observation 
period longer than $1/\Delta\omega$.
Therefore, data analyses using accurate emprical wave forms as templates
are necessary to enhance the sensitivity.
This point will be discussed elsewhere.

%
%
\section{Summary and Discussion.}

In this paper, we have discussed how to constrain the string 
axion parameters $(\mu, f_a)$ by
observations of continuous GWs from an axion cloud around a BH. 
The expected constraints from Cygnus X-1 are shown in 
Fig.~\ref{mass-fa-constraint-CygX1-aLIGO} for the LIGO and aLIGO detectors. 
A targeted search for continuous waves from Cygnus X-1 
is required in order to detect the signal
or derive the upper bound on the GW amplitude 
in the condition~\eqref{Condition1-CygX1}.
Such a targeted search should be possible, since similar analyses 
have been done already for 
neutron stars in binary systems \cite{Abbott:2006,Abbott:2007}.

There are other solar-mass BHs in binaries
for which the system parameters have been observationally studied,
and similar constraints can be obtained from these BHs. 
But we have to be careful about the BH spin parameters $a/M$, as they take various
values from zero to unity (see Table 1 of Ref.~\cite{Fragos:2014}). 
If we use a moderately spinning BH, a constraint can be imposed  
in a smaller range of the axion mass compared to the rapidly spinning case, 
because an axion cloud in the $\ell = m = 1$ mode forms only when the 
superradiant condition $\mu\approx \omega\le \Omega_H$ is satisfied.
For the spin parameter $a/M=0.7$, e.g., 
the  $\ell=m=1$ mode is unstable for $M\mu\lesssim 0.2$, and hence, 
the constraint can be discussed only in 
the range $\mu\lesssim 1.6\times 10^{-12}\mathrm{eV}$
for the BH mass $M=15M_\odot$. 
Except for this point, the same method holds as well. 
Although the growth rate of the superradiant instability 
becomes smaller as $a/M$ is decreased, 
this is not a problem because its time scale is still much shorter
than the age of the BH and there is enough time for the formation
of an axion cloud. Also, the wave form of continuous GWs 
scarcely changes with the value of the spin parameter.

Finally, we discuss the possiblity of
detecting continuous GWs from an axion cloud
around an {\it isolated} BH. In addition to 
visible BHs, $10^8$--$10^9$ 
isolated BHs are expected to exist in our galaxy \cite{Brown:1993,Timmes:1995}.
Since the averaged distance between two neighboring BHs is 
$\sim 10$ pc in this estimate, the detection
may be easier compared to the case of Cygnus X-1.
But in this case, 
we have to explore the method to distinguish
GWs from an axion cloud 
and those from a distorted neutron star, since 
the recent theoretical studies suggest that neutron stars could generate 
detectable GWs as 
well~\cite{Horowitz:2009,JohnsonMcDaniel:2012,Johnson-McDaniel:2013}. 
For this purpose,  
more detailed modeling of wave forms 
including the axion self-interaction effect
is necessary.  
The frequency modulation due to radial oscillation
of an axion cloud should provide a smoking gun for 
GWs of axion origin.

\ack

We thank Keith Riles for discussions and various suggestions.
We thank the Yukawa Institute for Theoretical Physics
at Kyoto University for hospitality during the YITP-T-14-1 
workshop on ``Holographic vistas on Gravity and
Strings,'' where part of this work has been done.
This work was 
supported by the Grant-in-Aid for Scientific Research (A) 
(Numbers 22244030 and 26247042)
from Japan Society for the Promotion of Science (JSPS).


%

\end{document}